\documentstyle[12pt,epsf]{article}

\catcode`\@=11
\long\def\@makefntext#1{ 
\protect\noindent \hbox to 3.2pt {\hskip-.9pt  
$^{{\ninerm\@thefnmark}}$\hfil}#1\hfill} 

\def\thefootnote{\fnsymbol{footnote}}
 \def\@makefnmark{\hbox to 0pt{$^{\@thefnmark}$\hss}}  
	
\def\ps@myheadings{\let\@mkboth\@gobbletwo
\def\@oddhead{\hbox{} 
\rightmark\hfil\ninerm\thepage}   
\def\@oddfoot{}\def\@evenhead{\ninerm\thepage\hfil 
\leftmark\hbox{}}\def\@evenfoot{}
\def\sectionmark##1{}\def\subsectionmark##1{}}

\begin{document}

\newcommand{\symbolfootnote}{\renewcommand{\thefootnote}
	{\fnsymbol{footnote}}}
\renewcommand{\thefootnote}{\fnsymbol{footnote}}
\newcommand{\alphfootnote}
	{\setcounter{footnote}{0}
	 \renewcommand{\thefootnote}{\sevenrm\alph{footnote}}}

\newcounter{sectionc}\newcounter{subsectionc}\newcounter{subsubsectionc}
\renewcommand{\section}[1] {\vspace{0.6cm}\addtocounter{sectionc}{1} 
\setcounter{subsectionc}{0}\setcounter{subsubsectionc}{0}\noindent 
	{\bf\thesectionc. #1}\par\vspace{0.4cm}}
\renewcommand{\subsection}[1] {\vspace{0.6cm}\addtocounter{subsectionc}{1} 
	\setcounter{subsubsectionc}{0}\noindent 
	{\it\thesectionc.\thesubsectionc. #1}\par\vspace{0.4cm}}
\renewcommand{\subsubsection}[1] {\vspace{0.6cm}\addtocounter{subsubsectionc}{1}
	\noindent {\rm\thesectionc.\thesubsectionc.\thesubsubsectionc. 
	#1}\par\vspace{0.4cm}}
\newcommand{\nonumsection}[1] {\vspace{0.6cm}\noindent{\bf #1}
	\par\vspace{0.4cm}}
					         
\newcounter{appendixc}
\newcounter{subappendixc}[appendixc]
\newcounter{subsubappendixc}[subappendixc]
\renewcommand{\thesubappendixc}{\Alph{appendixc}.\arabic{subappendixc}}
\renewcommand{\thesubsubappendixc}
	{\Alph{appendixc}.\arabic{subappendixc}.\arabic{subsubappendixc}}

\renewcommand{\appendix}[1] {\vspace{0.6cm}
        \refstepcounter{appendixc}
        \setcounter{figure}{0}
        \setcounter{table}{0}
        \setcounter{equation}{0}
        \renewcommand{\thefigure}{\Alph{appendixc}.\arabic{figure}}
        \renewcommand{\thetable}{\Alph{appendixc}.\arabic{table}}
        \renewcommand{\theappendixc}{\Alph{appendixc}}
        \renewcommand{\theequation}{\Alph{appendixc}.\arabic{equation}}
        \noindent{\bf Appendix \theappendixc #1}\par\vspace{0.4cm}}
\newcommand{\subappendix}[1] {\vspace{0.6cm}
        \refstepcounter{subappendixc}
        \noindent{\bf Appendix \thesubappendixc. #1}\par\vspace{0.4cm}}
\newcommand{\subsubappendix}[1] {\vspace{0.6cm}
        \refstepcounter{subsubappendixc}
        \noindent{\it Appendix \thesubsubappendixc. #1}
	\par\vspace{0.4cm}}

\def\abstracts#1{{
	\centering{\begin{minipage}{30pc}\tenrm\baselineskip=12pt\noindent
	\centerline{\tenrm ABSTRACT}\vspace{0.3cm}
	\parindent=0pt #1
	\end{minipage} }\par}} 

\newcommand{\bibit}{\it}
\newcommand{\bibbf}{\bf}
\renewenvironment{thebibliography}[1]
	{\begin{list}{\arabic{enumi}.}
	{\usecounter{enumi}\setlength{\parsep}{0pt}
\setlength{\leftmargin 1.25cm}{\rightmargin 0pt}
	 \setlength{\itemsep}{0pt} \settowidth
	{\labelwidth}{#1.}\sloppy}}{\end{list}}

\topsep=0in\parsep=0in\itemsep=0in
\parindent=1.5pc

\newcounter{itemlistc}
\newcounter{romanlistc}
\newcounter{alphlistc}
\newcounter{arabiclistc}
\newenvironment{itemlist}
    	{\setcounter{itemlistc}{0}
	 \begin{list}{$\bullet$}
	{\usecounter{itemlistc}
	 \setlength{\parsep}{0pt}
	 \setlength{\itemsep}{0pt}}}{\end{list}}

\newenvironment{romanlist}
	{\setcounter{romanlistc}{0}
	 \begin{list}{$($\roman{romanlistc}$)$}
	{\usecounter{romanlistc}
	 \setlength{\parsep}{0pt}
	 \setlength{\itemsep}{0pt}}}{\end{list}}

\newenvironment{alphlist}
	{\setcounter{alphlistc}{0}
	 \begin{list}{$($\alph{alphlistc}$)$}
	{\usecounter{alphlistc}
	 \setlength{\parsep}{0pt}
	 \setlength{\itemsep}{0pt}}}{\end{list}}

\newenvironment{arabiclist}
	{\setcounter{arabiclistc}{0}
	 \begin{list}{\arabic{arabiclistc}}
	{\usecounter{arabiclistc}
	 \setlength{\parsep}{0pt}
	 \setlength{\itemsep}{0pt}}}{\end{list}}

\newcommand{\fcaption}[1]{
        \refstepcounter{figure}
        \setbox\@tempboxa = \hbox{\tenrm Fig.~\thefigure. #1}
        \ifdim \wd\@tempboxa > 6in
           {\begin{center}
        \parbox{6in}{\tenrm\baselineskip=12pt Fig.~\thefigure. #1 }
            \end{center}}
        \else
             {\begin{center}
             {\tenrm Fig.~\thefigure. #1}
              \end{center}}
        \fi}

\newcommand{\tcaption}[1]{
        \refstepcounter{table}
        \setbox\@tempboxa = \hbox{\tenrm Table~\thetable. #1}
        \ifdim \wd\@tempboxa > 6in
           {\begin{center}
        \parbox{6in}{\tenrm\baselineskip=12pt Table~\thetable. #1 }
            \end{center}}
        \else
             {\begin{center}
             {\tenrm Table~\thetable. #1}
              \end{center}}
        \fi}

\def\@citex[#1]#2{\if@filesw\immediate\write\@auxout
	{\string\citation{#2}}\fi
\def\@citea{}\@cite{\@for\@citeb:=#2\do
	{\@citea\def\@citea{,}\@ifundefined
	{b@\@citeb}{{\bf ?}\@warning
	{Citation `\@citeb' on page \thepage \space undefined}}
	{\csname b@\@citeb\endcsname}}}{#1}}

\newif\if@cghi
\def\cite{\@cghitrue\@ifnextchar [{\@tempswatrue
	\@citex}{\@tempswafalse\@citex[]}}
\def\citelow{\@cghifalse\@ifnextchar [{\@tempswatrue
	\@citex}{\@tempswafalse\@citex[]}}
\def\@cite#1#2{{$\null^{#1}$\if@tempswa\typeout
	{IJCGA warning: optional citation argument 
	ignored: `#2'} \fi}}
\newcommand{\citeup}{\cite}

\def\fnm#1{$^{\mbox{\scriptsize #1}}$}
\def\fnt#1#2{\footnotetext{\kern-.3em
	{$^{\mbox{\sevenrm #1}}$}{#2}}}

\font\twelvebf=cmbx10 scaled\magstep 1
\font\twelverm=cmr10 scaled\magstep 1
\font\twelveit=cmti10 scaled\magstep 1
\font\elevenbfit=cmbxti10 scaled\magstephalf
\font\elevenbf=cmbx10 scaled\magstephalf
\font\elevenrm=cmr10 scaled\magstephalf
\font\elevenit=cmti10 scaled\magstephalf
\font\bfit=cmbxti10
\font\tenbf=cmbx10
\font\tenrm=cmr10
\font\tenit=cmti10
\font\ninebf=cmbx9
\font\ninerm=cmr9
\font\nineit=cmti9
\font\eightbf=cmbx8
\font\eightrm=cmr8
\font\eightit=cmti8


\centerline{\tenbf FINITE TEMPERATURE EFFECTIVE POTENTIAL FOR  }
\baselineskip=16pt
\centerline{\tenbf 
SPONTANEOUSLY BROKEN $\lambda \Phi^4$ THEORY.\footnote{
Talk held at the {\it Workshop on Quantum Infrared Physics}, 6-10 June
1994 at Paris}
}
\vspace{0.8cm}
\centerline{\tenrm HERBERT NACHBAGAUER
\footnote{email: herby @ lapphp8.in2p3.fr}}
\baselineskip=13pt
\centerline{\tenit 
Laboratoire de Physique Th\'eorique ENSLAPP }
\baselineskip=12pt
\centerline{\tenit 
Chemin de Bellevue, BP 110, F--74941 Annecy-le-Vieux Cedex, France }
\vspace{0.3cm}
\abstracts{
We present a self-consistent calculation of the finite temperature
effective potential for $\lambda \Phi^4$ theory in four dimensions
using a composite operator effective action. We find that in a
spontaneously broken theory not only the so-called daisy and
superdaisy graphs contribute to the next-to-leading order thermal
mass,   but
also resummed non-local diagrams are of the same order, thus
altering the effective potential at small effective
mass.
}

\vfil
\twelverm   
\baselineskip=14pt
\section{Introduction }
The study of $\lambda \Phi^4$-theory at finite temperature 
is of great interest for a wide field
of applications. A self-interacting  scalar field 
serving as a simple model for the Higgs particle in the standard  model of  
electroweak interactions may allow the study of  symmetry changing phase
transitions. In fact, the order of the electroweak phase transition
plays a crucial r\^ole  
in the framework of cosmological scenarios  as well  as  for
the badly understood process of baryogenesis\cite{kolbt}.
Despite the simplicity of the model, one may at least hope to gain
some insight in the mechanism  of the phase transition. Moreover, the
theory is a suitable test ground for analytic non-perturbative
methods, e.g. variational methods\cite{stev} as well as for lattice
simulations.

High temperature symmetry restoration in a
spontaneously broken theory was already noted by Kirzhnits and 
Linde\cite{kirzl}
and worked out quantitatively subsequently\cite{dwk}.
The convenient tool for studying the behavior of the theory turns out
to be the effective potential. Whereas the approximate 
critical temperature is
already determined by the one loop potential, the order of
the phase transition depends
on the detailed shape of the potential requiring  also 
the analysis of
higher loop contributions, even for small coupling constant.

In particular, one finds quantizing the theory around the classical
non-trivial minimum that the one loop self-energy at high temperature
behaves like $m_T \sim \sqrt{\lambda} T $, independent of external 
momenta and the value of the classical minimum. Thus, for large $T$
the thermal
mass dominates over the tree-level mass and the minimum of the 
effective action becomes the trivial one.
The theory exhibits two important features: non-temperature stable vacuum
and effective temperature dependent mass. It was consequently
proposed\cite{stev}  to move
the tree level mass into the interaction part of the Lagrangian and to
start perturbation theory with the free Lagrangian 
\mbox{ ${1 \over 2}  \Phi (\Box + \Omega^2)\Phi $ }
including a yet undetermined mass  parameter $\Omega$. Then one 
calculates the effective  potential and fixes the parameter by the 'principle 
of minimal sensitivity' $\partial V(\Phi_{min}) / \partial \Omega=0$.
 
A more systematic approach of a self-consistent loop expansion was developed
already some time ago by Cornwall, Jackiw and Tomboulis (CJT) in their 
effective action formalism for composite operators\cite{CJT}. The basic idea 
is to  introduce a bilocal mass operator instead of the 
local mass $\Omega$
in the generating 
functional. Then one defines a generalized effective action as the double
Legendre transform of the generating functional,
  which  is now not only a 
functional of the expectation value of the field,
$\phi=\left< \Phi \right> $ but also depends on the 
expectation value of the time-ordered product $\mbox{T} \Phi(x) \Phi(y)$.
The principle of minimal sensitivity gets replaced by the functional 
condition
$\delta \Gamma (\phi,G) / \delta G(x,y) = 0 $ which is employed to solve for the
unknown exact two-point Green function $G_0(\phi;x,y)$, and the conventional 
effective action can be reconstructed as  $\Gamma (\phi,G_0(\phi;x,y) )$. 
For a given action $I(\Phi )$ one finds\cite{CJT} 
\def\tr{\mbox{Tr}}
$$ \Gamma (\phi,G) = I(\phi) +{1 \over 2 } \tr \ln 
( D_0 G^{-1}  )
+{1 \over 2 } \tr ( D^{-1} G - {\bf 1} ) + \Gamma^{(2)} (\phi,G) ,$$
$$ D^{-1 }(\phi;x,y) = { \delta^2 I(\phi )\over \delta \phi(x) 
\delta \phi(y) } ,$$ 
where $ D_0 (x,y )$ is  the free propagator derived from the part of 
the action which is quadratic in the fields. 
The quantity $ \Gamma^{(2)} (\phi,G) $ contains all two--loop
contributions and higher and has to be calculated as follows: 
Shift the field $\Phi$ in the classical action by $\phi$.
Then $I(\Phi + \phi )$ contains terms cubic and higher in $\Phi$
which define the vertices. $\Gamma^{(2)} (\phi,G) $ contains  
all two-particle-irreducible vacuum graphs with 
the given set of vertices and the propagators replaced by 
$G(x,y)$. The gap equation now reads
$$ G^{-1} (x,y) = D^{-1} (x,y) + 2 {\delta \Gamma^{(2)} (\phi,
G) \over \delta G(x,y ) } .$$

\section{ $\lambda \Phi^4 $ - theory }

The graphs naively contributing up to order 
$\lambda^2  $ are shown in Fig. 1.
\begin{figure}
\hspace*{\fill}
\epsfxsize3in
\epsfbox{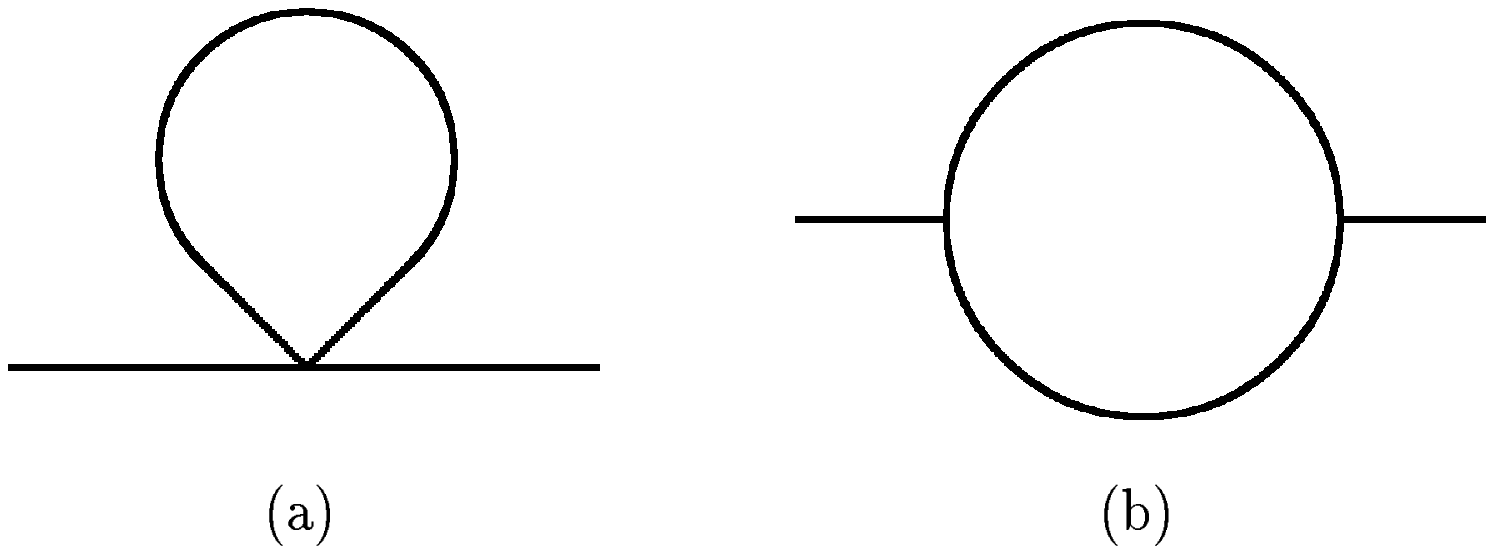}
\hspace*{\fill}\vspace{2em}
\fcaption{Feynman graphs contributing to lowest order to the self
energy. Lines denote the function $G(x,y)$, the three-vertex is
given by $- \lambda \phi / 3! $ and the four-vertex by $- \lambda /4! $.
} 
\end{figure} 
Although the non-local
contribution  Fig.~1b is 
formally of order $\lambda^2$ it contributes to order $\lambda$
in the false vacuum since already
the tree-level
value of the expectation value of the field is 
$ \phi \sim m/\sqrt{\lambda} $. Consequently, the 3-vertex given
by $\lambda \phi $ is of order $\sqrt{\lambda} $ which means that 
in the case of a spontaneously broken theory 
the non-local contribution (b) is as important as the local one (a).
We will show that this is also true in the quantized theory.
Thus, the Hartree--Fock approximation involving only the local 
graph which corresponds  to the superdaisy or hard thermal loop resummation
at finite temperature fails to be consistent.
We therefore  include the non-local graph Fig. 1b in the
self energy and the gap equation in Fourier-space reads  \goodbreak
\begin{eqnarray} 
 G^{-1} (\vec k,\omega)=\vec k^2 + \omega^2 + 
(-m^2 + {\lambda \over 2}
\phi^2 ) + {\lambda \over 2 } T \sum_{\omega '} \int {d^3 p \over 
(2 \pi)^3 } G(\omega ',\vec p )  + \hspace*{1cm} 
\nonumber \\
 + { 1 \over 2 } ( \lambda \phi)^2  T 
\sum_{\omega '} \int {d^3 p \over (2 \pi)^3 }
G(\omega,\vec p) G(\omega + \omega ',\vec p + \vec k ) 
\label{5} 
\end{eqnarray} 
where we have written down the expression in imaginary time formalism.
The first term corresponding to graph (a) does not depend on external momenta
and energies and thus can be treated as a mass term. A solution
for the gap equation involving only the constant term can easily be
found\cite{pi} and corresponds to superdaisy resummation.
However, we are interested in the behavior of a broken theory, where the last 
term in Eq. (\ref{5}) is of the same order.
We employ  the Ansatz
$$ G^{-1} (\omega,\vec k) = \vec k^2 + \omega^2 + \mu^2 + \Pi(\omega,\vec k) 
$$ 
and subsequently absorb all local contributions in a mass parameter 
$\mu$ which will be determined self-consistently afterwards.
Plugging the Ansatz into Eq. (\ref{5}) recasts the gap equation into 
an equation for $\Pi(\omega,\vec k)$.
Simple power counting now reveals that the 
second term in Eq. (\ref{5})  diverges  for large momenta $\vec p$. 
However, one may absorb the divergent part in a mass renormalization  
rendering  the self energy $\Pi ( \omega ,\vec k) $  a finite
quantity. 

Since we are only interested in the high temperature ($T \gg \mu $)  regime
 of the theory one may further simplify Eq. (\ref{5}) by taking only the 
 $\omega ' =0$ mode in the second sum, since all other terms are at least 
 suppressed by a factor $1 / T $. This in turn means that it is also only the zero
mode of the self energy, that contributes most. Putting $\Pi( k) =
\Pi( \omega=0,|\vec k |) $ we end up with the integral equation 
\begin{eqnarray} 
\lefteqn{   \Pi( k ) =  } \nonumber \\
& & 
{T \over 2 } ( \lambda \phi )^2 \int\limits_0^\infty {dp p^2 
 \over (2 \pi )^2 } 
\int\limits_{-1}^1 {dz \over ( p^2 + \mu^2 + \Pi( p ) )
( p^2 + k^2 
+ 2 p k z + \mu^2 + \Pi (\sqrt{p^2 + k^2 + 2 p k z } ) ) } \nonumber \\
&&   
\label{7} 
\end{eqnarray}  
which we are now going to solve. \newpage

\section{Solution of the gap equation and effective potential}

First,  from Eq. (\ref{7}) one immediately reads off the high 
momentum behavior $\Pi(k \to \infty ) =0$.
Secondly, we note that  due to
the non-local character of the integral equation we have to know
the function $\Pi (k) $ for all momenta,
 even if we were only interested in the 
low momentum behavior of the self energy. 
However, numerical study of this equation reveals that $\Pi (k)$
can be approximated by  
\begin{equation} 
\Pi(k) = \mu^2 {A(\kappa )^2 \over a(\kappa )^2 + (k/\mu)^2 },
\quad  \kappa = {T (\lambda \phi )^2 \over 8 \pi^2 \mu^3 } 
\label{8} \end{equation} 
where we introduced the dimensionless quantity $\kappa$.
Since $\Pi (k) $ is of relevance in the denominator of 
Eq. (\ref{7}) only in the infrared region $k \ll \mu $, we choose  
$A(\kappa)$ and $a(\kappa )$ such that Eq. (\ref{7}) is fulfilled exactly
for $\Pi(0)$ and $\Pi''(0)$. $A(\kappa )$ and $a(\kappa )$ 
are then determined by  
two coupled complicated transcendental equations\cite{myself}.
We note that the Ansatz (\ref{8}) may be interpreted as 
Yukawa-like potential with  screening mass $\mu a$ and strength $\sim A^2 $

Putting all together one finally gets the gap equation for the 
mass parameter $\mu$,
\begin{equation} 
 -m^2 + {\lambda \over 2 }\phi^2 + {\lambda \over 2} ( { T^2 \over 12 }
+ {T \mu \over 8 \pi} f(\kappa) ) = \mu^2 
\label{9} 
\end{equation} 
where we have absorbed the divergencies in redefined quantities $m,\; \lambda$ 
and $f=f(A(\kappa),a(\kappa) ) $. 
We find $f( 0 ) = -2 $ which corresponds to the value 
if one only had included the local contribution of graph (a). On the other hand
for large 
values of the parameter $\kappa$, which in the non-trivial vacuum  $\phi 
\ne 0$ means
small mass parameter $\mu$, one finds the asymptotic behavior
$f(\kappa ) \sim C \kappa^{1\over 3 } $ with $C \sim -1.895\ldots $.
Remarkably enough, the mass parameter cancels out in the next to leading
order contribution in Eq. (\ref{9}) and we are left with a term
 $T \mu f(\kappa ) \sim T^{4\over 3} (\lambda \phi )^{2\over 3}  $.
In view of the result, 
this quite surprising behavior can already estimated from 
Eq. (\ref{7}). Rescaling  by $\mu$ such that the self--energy becomes 
a dimensionless,  the equation
may roughly be read as $\Pi^3 \sim \kappa $.

We thus find that the non-local term included in this investigation 
significantly alters the gap equation (\ref{9}) at vanishing mass
$\mu $, 
that means 
when perturbation theory becomes problematic and infrared effects 
enter the game. 

Since, to lowest order, the effective mass is just the second derivative of 
the effective potential with respect to $\phi$, we are now able to reconstruct
the lowest order effective potential just by twice integrating the gap 
equation over to $\phi$. 
Keeping only the asymptotic form of $f$ in the gap equation,
different temperature regions may distinguished by their limiting values
$T_i = m \sqrt{24 \over \lambda } (1- F_i \lambda )^{-{1\over 2}}  $,
 where 
$F_i = ( 0 , {216\over 20 \sqrt{10} } |C|^{3/2} ,
{24 \sqrt{3} \over 5 \sqrt{5} } |C|^{3/2} ,
24\cdot 3^{-{3/2} } |C|^{3/2} ) $. 
The potential has one non-trivial minimum for $T<T_0$
and develops another for $T_0 < T < T_1 $. 
At $T=T_1 $ the trivial and the non-trivial minimum have equal depth.
As temperature is further raised until $ T = T_2$, the non-trivial minimum 
vanishes completely. There exists however still a value $\phi \neq 0$, where
 the bending
of the potential changes sign as long as the temperature is 
lower than $T_3$. 
The potential for $\lambda =0.1 $  is drawn in Fig. 2 for the corresponding
values of the limiting  temperatures where in addition, we added a curve for a 
temperature between $T_0 < T < T_1 $.

\section{Conclusion} 
We have shown, to lowest order however, that the naively expected 
resummed effective potential in a spontaneously broken theory fails to
be consistent. Instead we encountered an additional contribution 
to the effective mass 
in the gap equation behaving like $\phi^{2/3} $ which cannot 
be found in a perturbative calculation. This may point towards
a breakdown of the resummed perturbative calculation 
thus encouraging the extensive 
study of 
alternative methods.
Nevertheless, we find that the phase transition is still of first order.
A refinement of the present approach 
is currently under investigation. 



\begin{figure}
\epsfxsize4in 
\vspace*{-3cm} \hspace*{\fill} 
\epsfbox{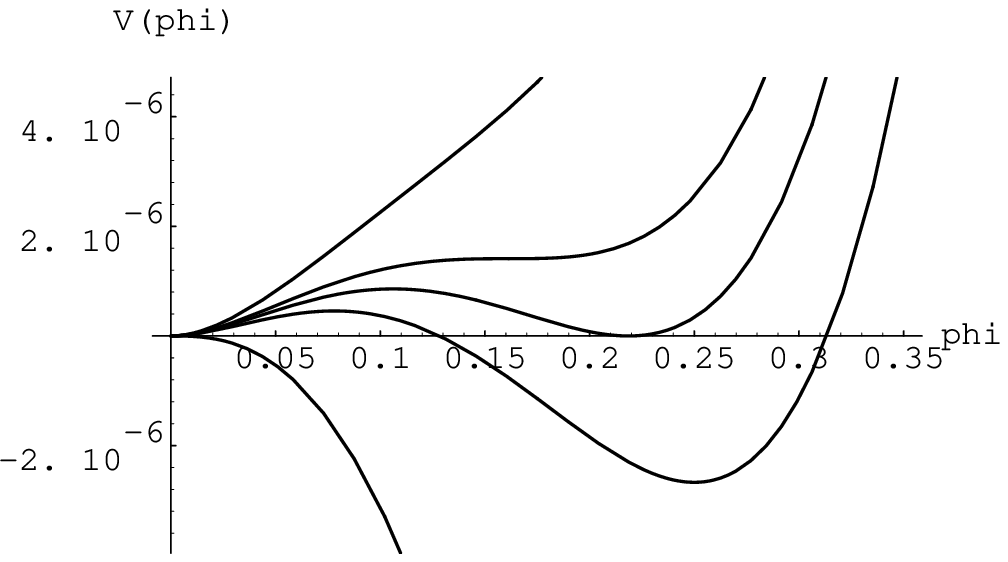}
\hspace*{\fill} 
\vspace*{-2cm} 
\fcaption{The finite temperature effective potential for $\lambda=0.1$ in 
units of the redefined mass $m$. The upper three curves correspond to the 
temperatures $T_3,T_2,T_1$ respectively, the lowest curve to the temperature
$T_0$. }
\end{figure} 
\hspace*{\fill}



\end{document}